\documentclass[twocolumn,preprintnumbers,amsmath,amssymb]{revtex4}
\usepackage{epsfig}
\begin{document}

\title{Ferromagnetic resonance study of polycrystalline Cobalt ultrathin films}
\author{J-M. L. Beaujour, W. Chen,  A. D. Kent} \affiliation{Department of Physics, New York University, 4
Washington Place, New York, NY 10003, USA}
\author{J. Z. Sun}
\affiliation{IBM T. J. Watson Research Center, Yorktown Heights,
NY 10598, USA}
\date{\today}

\begin{abstract}
We present room temperature ferromagnetic resonance (FMR) studies
of polycrystalline $||$Pt/10 nm Cu/\textit{t} Co/10 nm Cu/Pt$||$
films as a function of Co layer thickness (1 $\leq t\leq 10$ nm)
grown by evaporation and magnetron sputtering. FMR was studied
with a high frequency broadband coplanar waveguide (up to 25 GHz)
using a flip-chip method. The resonance field and the linewidth
were measured as a function of the ferromagnetic layer thickness.
The evaporated films exhibit a lower magnetization density
($M_{\text{s}}$ = 1131 emu/$\hbox{cm}^{3}$) compared to the
sputtered films ($M_{\text{s}}= 1333$ emu/$\text{cm}^{3}$), with
practically equal perpendicular surface anisotropy ($K_{\text{s}}
\simeq $ -0.5 erg/$\text{cm}^{2}$). For both series of films, a
strong increase of the linewidth was observed for Co layer
thickness below 3 nm. For films with a ferromagnetic layer thinner
than 4 nm, the damping of the sputtered films is larger than that
of the evaporated films. The thickness dependence of the linewidth
can be understood in term of the spin pumping effect, from which
the interface spin mixing conductance
${\textrm{g}}^{\uparrow\downarrow}S^{-1}$ is deduced.
\end{abstract}

 \maketitle
\section{Introduction}
Polycrystalline ferromagnetic films a few nanometers thick are
commonly used in spin-transfer devices and, in general, in
spintronic applications. For instance, the spin torque effect is
typically studied in devices that consist of two magnetic layers
separated by a normal metal (NM) and two contact layers made of
the same NM on top and bottom of the structure \cite{kent1}. One
of the magnetic layers, about 10 nm thick, provides a spin
polarized current that is used to excite and switch the
magnetization of a very thin adjacent layer (few nm). The
threshold current density for magnetic excitations is proportional
to the Gilbert damping constant G, and the effective magnetization
\cite{sun}. In order to understand the physics of spin transfer,
it is therefore important to characterize the magnetic properties
and magnetic relaxation of ultrathin films. FMR is a sensitive
technique to study magnetic ultrathin films. It provides
information on the magnetization density, the magnetic anisotropy
and the damping. The precession of the magnetization $\vec{M}$
about an effective field $\vec{H}_{\text{eff}}$ is described by
the Landau-Lifshitz equation. For a polycrystalline film that is
magnetically saturated in the film plane, the resonance condition
is \cite{mills}:
\begin{equation}
\label{eq1} {\left({2 \pi f
\over{\gamma}}\right)^2}=H_{\text{res}}\ (H_{\text{res}}+4\pi
M_{\text{eff}})\;,
\end{equation}
where $\gamma=\textrm{g} \mu _B / \hbar$ is proportional to g, the
Land\'{e} g-factor. The effective field $4\pi$M$_{\text{eff}}$ is
defined as \cite{silva1} :
\begin{equation} \label{eq2}
{4\pi M_{\text{eff}}}=4\pi M_{\text{s}}+{2\
K_{\text{s}}\over{M_{\text{s}}\ t}}\;.
\end{equation}
The last term of Eq.~\ref{eq2} is the surface anisotropy field. It
characterizes a thickness dependent anisotropy associated with
interface anisotropy and/or strain-magnetoelastic interactions.
Another parameter of importance in FMR is the linewidth. The full
width at half power $\triangle H$ is commonly fitted to
\cite{mills}:
\begin{equation}
\label{eq3} {\triangle H}=\triangle H_0+{2\
\text{G}\over{{\gamma}^2 M_{\text{s}}}}2 \pi f\;.
\end{equation}
The constant $\triangle H_0$ is a phenomenological term related to
inhomogeneous broadening of the FMR line. The slope of $\triangle
H \  vs. \ f$ is directly proportional to the Gilbert damping
constant. The two terms, $\triangle H_0$ and the slope, are
referred to as the extrinsic and intrinsic contribution to the
linewidth respectively.

In this paper, we compare the FMR response of ferromagnetic
ultrathin films grown by evaporation and by sputtering. We first
discuss the sample fabrication and the experimental set up. Then
the thickness dependence of $4\pi M_{\text{eff}}$ and G will be
presented and analyzed.
\section{Experimental technique}
\begin{figure}
\begin{center}\includegraphics[width=6cm]{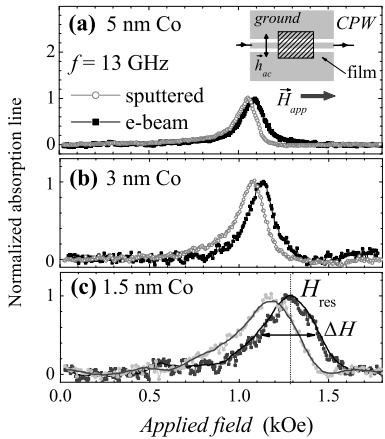}
\vspace{-4 mm}\caption{\label{fig1}The normalized absorption curve
at 13 GHz for a selection of Pt/Cu/Co/Cu/Pt films, where the
magnetic layer is (a) 5 nm, (b) 3 nm and (c) 1.5 nm thick. The
inset of (a) shows the experimental geometry.} \vspace{0 mm}
\end{center}
\end{figure}

The samples are made of a single polycrystalline Co layer embedded
between two Pt/Cu bilayers. Two series of samples were fabricated
by evaporation and sputtering. The samples were prepared in a UHV
system with a base pressure of $5\times10^{-8}$ Torr on polished
 semi-insulating GaAs (001) substrate of 350
 $\mu$m thickness. For the evaporated films, an e-beam was used to
evaporate the Pt layers and the Co layers. The Cu layers were
deposited using thermal evaporation. The second set of films were
made using magnetron sputtering. Those films have a thicker Pt
layer (5 nm) than the evaporated films (1.5 nm). In both series of
samples, the ferromagnetic layer thickness varied from 1 nm to 10
nm, while the Cu layer thickness was kept fixed at 10 nm. The FMR
measurements were carried out at room temperature using a coplanar
waveguide (CPW), designed to have a 50 $\Omega$ impedance within a
broad frequency range (up to 25 GHz). The device was fabricated on
a similar GaAs wafer than the films employing a bi-layer
photoresist. The metallic layer is made of 1.5 nm Pt for adhesion,
and 200 nm Au. The waveguide, 4 mm long, has a transmission line
of 50 $\mu$m width and a gap to the ground lines of 32 $\mu$m. The
two ends of the line were directly connected to the ports of an
Agilent Network Analyzer. The CPW was employed as an ac magnetic
field generator with the assumption that the dominant mode was the
TEM mode, and as an inductive sensor. Samples were placed directly
on top of the CPW (flip-chip), as shown in the inset of
Fig.~\ref{fig1}a. A dc magnetic field (up to 4.5 kOe), generated
by an electromagnet was directed along the axis of the
transmission line and perpendicular to the ac magnetic field. The
dc applied field was measured with a Hall probe sensor, that was
calibrated using EPR on dpph, a spin 1/2 system. All the
measurements were done with the dc field and the ac field aligned
in the plane of the film. The absorption line from 4 GHz to 25 GHz
was measured by monitoring the relative change in the transmitted
power as a function of the applied magnetic field. As will be
shown below, the technique is sensitive enough to enable FMR
studies in Co magnetic layers as thin as 1 nm.

\section{Results}
Typical absorption lines at 13 GHz of a selection of
Pt/Cu/Co/Cu/Pt evaporated and sputtered films are shown in
Fig.~\ref{fig1}. The normalized data were obtained by subtracting
the background signal and dividing by the relative change in power
at resonance. The absorption lines are Lorentzians, with a slight
asymmetry observed at certain frequencies. For each film, the
effective magnetic field $4\pi M_{\text{eff}}$ and the g-factor
was deduced from the best fit of the experimental data,
$f^2/H_{\text{res}}$ $vs.$ $H_{\text{res}}$, to Eq.~\ref{eq1}. For
the two series, the g-factor does not exhibit a clear thickness
dependence. Within the error bars, the average values of g for the
evaporated and sputtered films are practically equal, $2.49 \pm
0.14$ and $2.36 \pm 0.06$, respectively. Nevertheless, this is
about $25\%$ larger than the value reported in the literature for
fcc Co films (g$=2.14$) \cite{wiedwald}.
\begin{figure}
\begin{center}\includegraphics[width=5cm]{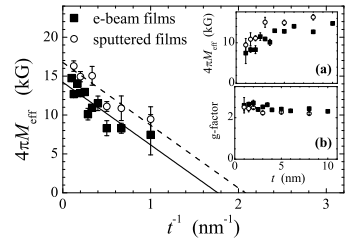}
\vspace{-2 mm}\caption{\label{fig2}Effective field versus inverse
 thickness for sputtered and evaporated
films. The solid and dash line are fits of the experimental data
based on Eq.~\ref{eq2}. The insets (a) and (b) shows $4\pi
M_{\text{eff}}$ and the g-factor versus thickness, respectively.}
\vspace{-6 mm}
\end{center}
\end{figure}
The effective fields exhibit a clear thickness dependence:
decreasing about 6 kOe when the Co layer thickness is decreased
from 10 nm to 1 nm (Fig.~\ref{fig2}a). The best fit to the
experimental data with Eq.~\ref{eq3} gives the surface anisotropy
constant for e-beam films $K_{\text{s}}=(-0.46\pm0.04)\
$erg/cm$^{2}$ and for sputtered films
$K_{\text{s}}=(-0.54\pm0.12)\ $erg/cm$^{2}$. The negative sign
reflects a perpendicular magnetic surface anisotropy. Within the
error bar, the surface anisotropy is practically independent of
the film deposition technique. In contrast, the sputtered films
exhibit a larger magnetization density, $M_{\text{s}}=1333\
$emu/cm$^{3}$, compared to the films prepared by evaporation
$M_{\text{s}}=1131\ $emu/cm$^{3}$. Those values are smaller than
the bulk magnetization density of fcc Cobalt ~1400 emu/cm$^{3}$.
The results are in good agreement with previous work conducted on
$t$ Co/2.5 nm Cu (111) epitaxial superlattices ($0.5 \leq t \leq
4$ nm) grown on GaAs(110) wafer \cite{O'Handley} . Indeed, it was
found that those films exhibit a surface anisotropy constant
$K_{\text{s}}=-0.47$ erg/cm$^{2}$ and have a density of
magnetization of 1241 emu/cm$^{3}$ on average.

\begin{figure}
\begin{center}\includegraphics[width=8cm]{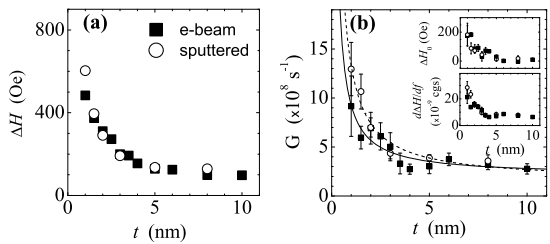}
\vspace{-3 mm}\caption{\label{fig3} (a) Thickness dependence of
the full width at half power, $\triangle H$, at 14 GHz. (b) The
Gilbert damping versus the Co layer thickness $t$, deduced from
the slope $d\triangle H/df$. The top inset shows the intercept
$\triangle H_0$ versus thickness. The lower inset shows the slope
$d \triangle H/ df$ versus thickness.}\vspace{-6 mm}
\end{center}
\end{figure}

The full width at half power $\triangle H$ was also studied as a
function of the frequency. The linewidth increases linearly with
frequency from 10 GHz to 25 GHz for both evaporated and sputtered
films. Fig.~\ref{fig3}a shows the thickness dependence of
$\triangle H$ for data recorded at 14 GHz. The linewidth is
significantly enhanced for films with Co thickness below 5 nm.
Following Eq.~\ref{eq3}, the parameters $\triangle H_{0}$ and the
slope $d\triangle H/df$ were extracted. The two contributions to
the linewidth exhibit similar thickness dependence, characterized
by a strong increase for Co layers thinner than 5 nm (inset of
Fig.~\ref{fig3}a).
$\triangle H_0$ is close to 0 Oe for films with Co layer thicker
than 5 nm, and it reaches 200 Oe when the ferromagnetic layer is 1
nm thick. The slope is about constant for $t\geq$4 nm, and it
increases by a factor 3 for the thinnest film. The Gilbert damping
constant was estimated from Eq.~\ref{eq3} and its thickness
dependence is shown in Fig.~\ref{fig3}b. We used the average
g-factor and the magnetization of saturation obtained from the
study of the thickness dependence of the effective field. The
sputtered and evaporated films have equivalent damping constant
for thick Co layers. However, with decreasing FM layer thickness,
G increases more rapidly for sputtered films than for evaporated
films. The enhancement of the damping for the ultrathin
ferromagnetic films can be interpreted in terms of the spin
pumping effect. Brataas and co-workers recently proposed a
mechanism for additional Gilbert damping in NM/FM/NM structures,
where the NM layers are two perfect spin sinks \cite{tserkovnyak}.
In this case, the additional damping scales inversely with the
thickness of the FM layer:
\begin{equation}
\label{eq4}
\hbox{G}(t)=\hbox{G}_0+{{({\hbox{g}\mu_B})^2}\over{2\pi\hbar}}{{{\textrm{g}}^{\uparrow\downarrow}S^{-1}}\over{t}}\;,
\end{equation}
where G$_0$ is the bulk damping and g$^{\uparrow\downarrow}S^{-1}$
is the spin mixing conductance at the FM/NM interface. The best
fit was obtained for $\text{G}_0=(2.09\pm0.44)\times 10^8$
s$^{-1}$ and
g$^{\uparrow\downarrow}S^{-1}=(0.89\pm0.12)\times10^{15}$
cm$^{-2}$ for the evaporated films, and G$_0=(1.52\pm0.71)\times
10^8 $ s$^{-1}$ and
g$^{\uparrow\downarrow}S^{-1}=(1.63\pm0.18)\times10^{15}$
cm$^{-2}$ for the sputtered films. The method of deposition has
practically no effect on G$_0$. For the spin mixing conductance,
there is a difference in the result for sputtered and evaporated
films. The value of the spin mixing conductance of the sputtered
films is close to the value calculated for a clean Co/Cu
interface, reported in Ref.\cite{tserkovnyak}
(g$^{\uparrow\downarrow}S^{-1}=1.41\times10^{15}$ cm$^{-2}$). For
the evaporated films, g$^{\uparrow\downarrow}S^{-1}$ is about two
times smaller. Eq. \ref{eq4} is valid for a system where the NM
layers are perfect spin sinks, i.e. the angular momentum generated
by the magnetization in precession diffuses and  is dissipated in
the NM layers. Mizukami \textit{et al.} showed that the linewidth
of Cu/Py/Cu/Pt sputtered films decreases with Pt layer thickness
decreasing below 1.5 nm \cite{mizukami}. Therefore, the smaller
value of g$^{\uparrow\downarrow}S^{-1}$ found for the evaporated
films compared to the sputtered films could be explained by the
thin Pt layers that do not act as perfect spin sinks. The results
suggest that the spin diffusion length of Pt at room temperature
is around 2 nm. Another mechanism that gives rise to the
broadening of the linewidth with decreasing thickness is the
two-magnon scattering mechanism. Arias and Mills
 calculated that this contribution is proportional to
$H_{\text{s}} ^2$, where $H_{\text{s}}$ is the interface magnetic
anisotropy, and so scales as 1/$t^n$ with $n=2$ \cite{arias}. A
log-log plot of the thickness dependence of the linewidth at 14
GHz, for instance, for sputtered and evaporated films gives
$n\simeq 0.8\pm0.1$. Furthermore, we conducted FMR measurements on
films without Pt layers, and with Co thickness layer of 2 and 3
gffnm. The Gilbert damping is about the value found for the
thickest films with Pt layers (to be published). Therefore, the
two-magnon scattering mechanism does not appear relevant to
understanding the thickness dependence of $\triangle H$.
\section{conclusion}
Polycrystalline Pt/Cu/Co/Cu/Pt ultrathin films exhibit smaller
magnetization density compared to the bulk material. Futhermore,
sputtered films exhibit a larger magnetization of saturation
compared to the evaporated films. In contrast, the interface
anisotropy is not affected by the deposition technique. The
linewidth increases strongly with thickness decreasing below 4 nm.
The estimated Gilbert damping shows similar behaviour. The spin
mixing conductance at Co/Cu interface was calculated and found to
be smaller for evaporated films with a Pt layer of 1.5 nm than for
sputtered films that have a Pt layer of 5 nm.

This research is supported by NSF-DMR-0405620 and by ONR
N0014-02-1-0995.

\end{document}